\documentclass[11pt,a4paper]{article}
\usepackage{amsfonts}
\usepackage{amssymb}
\usepackage{jheppub}
\usepackage{ulem}
\usepackage{extarrows}
\usepackage{multirow}
\usepackage{float}

\usepackage{inputenc}
\usepackage{xspace}
\usepackage{url}

\newcommand{\bea}{\begin{eqnarray}}
\newcommand{\eea}{\end{eqnarray}}
\newcommand{\bean}{\begin{eqnarray*}}
\newcommand{\eean}{\end{eqnarray*}}

\def\Label#1{\label{#1}%
  \smash{\hbox to0pt{\raise1ex\hbox{\tiny[#1]}\hss}}}

\renewcommand{\eqref}[1]{eq.~(\ref{#1})}

\newcommand{\uniHH}{II. Institut f\"ur Theoretische Physik, Universit\"at Hamburg, \\
Luruper Chaussee 149, D- 22761 Hamburg, Germany}
\newcommand{\rminfn}{I.N.F.N. Sezione di Roma ``Tor Vergata", Via della Ricerca Scientifica, 00133 Roma, Italy}

\allowdisplaybreaks
\title{Recursion relations from soft theorems}
\author[a]{Hui Luo,}
\author[b]{\,Congkao Wen}
\affiliation[a]{\uniHH}
\affiliation[b]{\rminfn}

\emailAdd{hui.luo@desy.de}
\emailAdd{Congkao.Wen@roma2.infn.it}
\abstract{We establish a set of new on-shell recursion relations for amplitudes satisfying soft theorems. The recursion relations can apply to those amplitudes whose additional physical inputs from soft theorems are enough to overcome the bad large-$z$ behaviour. This work is a generalization of the recursion relations recently obtained by Cheung et al for amplitudes in scalar effective field theories with enhanced vanishing soft behaviours, which can be regarded as a special case of those with non-vanishing soft limits. We apply the recursion relations to tree-level amplitudes in various theories, including amplitudes in the Akulov-Volkov theory and amplitudes containing dilatons of spontaneously-broken conformal symmetry.  
 }
\begin{document}
\maketitle
\section{Introduction}

Britto-Cachazo-Feng-Witten (BCFW) recursion relations~\cite{Britto:2004nc, Britto:2005fq} have played a central role in the development of the modern S-matrix program. 
The original recursion relations for the gauge theories were later generalized for amplitudes in gravity, string theory, as well as planar loop integrands in $\mathcal{N}=4$ super Yang-Mills theories~\cite{Bedford:2005yy,Cachazo:2005ca, Boels:2010bv,Cheung:2010vn,ArkaniHamed:2010kv}. 
With recursion relations, one can determine all the amplitudes from the lowest-point amplitudes.
Typically, those lowest-point results are fixed by the symmetries of theories completely. 

We begin with briefly reviewing the ideas of BCFW recursion relations. 
The recursion relations can be derived from a complex deformation of a given $n$-point amplitude (denoted as $A_n$) by performing so-called BCFW momentum shifts on two external particles,
\bea \label{eq:usualBCFW}
{p}_{\hat{i }} = p_i + z q \,, \quad  {p}_{\hat{j }} = p_j - z q \, .
\eea
From now on, we denote the shifted momenta with hat. 
The auxiliary momentum $q$ satisfies the constraints $q^2=q\cdot p_i = q\cdot p_j=0$ to ensure the on-shell conditions of ${p}_{\hat{i }}$ and ${p}_{\hat{j }}$\footnote{It is easy to see that there is no non-trivial solution for $q$ in the case where the space-time dimensions are smaller than four. Generalized recursion relations for amplitudes in three dimensions were obtained in~\cite{Gang:2010gy}.}. 
Under the shifts, a tree-level amplitude becomes $A_n(z)$, a complex function of $z$.
Then the undeformed amplitude, $A_n=A_n(0)$, can be written as a contour integral,
\bea
A_n(0) ={1 \over 2 \pi i} \oint_{|z|=0} {dz \over z} A_n(z) \,.
\eea
Applying Cauchy's contour theorem and the fact that tree-level amplitudes only have simple poles from the factorizations, the amplitude $A_n(0)$ can be expressed in a recursive form in terms of lower-point amplitudes as well as a possible contribution of the pole at $z=\infty$, 
\bea
A_n(0) = \sum_I A_{L}(z_I)  {1 \over P_I^2 - m^2_I} A_{R}(z_I) + R_{\infty}\, .
\eea
Here we sum over all factorization channels $I$ with shifted legs on different sides of factorization diagrams. 
$A_{L}(z_I)$ and $A_{R}(z_I)$ are shifted lower-point amplitudes entering the diagrams, with $z_I$ fixed by the on-shell condition $P_I^2(z_I)= m_I^2$. 
For many theories such as gauge theories, gravity, $R_{\infty}=0$~\cite{ArkaniHamed:2008yf}, we then have the recursion relations, which allow to determine higher-point amplitudes from lower-point. 

However, there are still a wide range of interesting theories whose amplitudes have a bad large-$z$ behaviour (namely $R_{\infty} \neq 0$), in particular, effective field theories with higher-dimensional operators as we will discuss in this paper. 
The extension of BCFW recursion relations in these more general theories is not straightforward. 
The obstruction arises due to limitation of our knowledge about $R_{\infty}$ except only a few special cases (see for instance~\cite{Penante:2014sza}). 
There have been some attempts to deal with $R_{\infty}$~\cite{Feng:2011twa, Benincasa:2011pg,Feng:2011jxa, Jin:2014qya,Feng:2015qna, Jin:2015pua}. 
One particular idea is to use the zeroes of amplitudes as additional information (besides simple poles from factorizations) to determine the unknown boundary term at $z=\infty$, to complete the recursive construction of any higher-point tree-level amplitudes. 
This idea appeared first in~\cite{Benincasa:2011pg}, and was further studied in~\cite{Feng:2011jxa} later. However, unlike the factorization poles, it is usually complicated and difficult to determine the zeroes of a given amplitude. 
It would be certainly interesting to explore the general structures and the physics when amplitudes become zero.
In a particular case, for amplitudes vanishing in the single-soft limit, the soft theorems can in fact provide simple and useful information about zeroes of the amplitudes~\cite{Cheung:2015ota}, at least a subset of zeroes. 
With the help of soft theorems, new recursion relations are proposed for the S-matrices of various effective field theories of pure scalars~\cite{Cheung:2015ota} . 
To ensure the recursion relations practicable, it is required that the amplitudes should vanish fast enough compared with the speed of the growth of the amplitudes in the large-$z$ limit. 
It is named as ``enhanced soft behaviours", whose concrete requirement will become clear shortly in the review section about the recursion relations in~\cite{Cheung:2015ota}. 

The topic of soft theorems is a rich subject in the study of scattering amplitudes, which can be traced back to~\cite{Low:1958sn, Weinberg:1964ew, Adler:1964um, Weinberg:1965nx}. 
Recently there has been great renewed interest on soft theorems due to the work~\cite{Strominger:2013jfa, He:2014laa}, and many new and interesting results have been obtained. 
In many examples, the amplitudes have either non-vanishing soft limits, or the massless soft particles are not scalars. 
Such cases go beyond what have been studied in~\cite{Cheung:2015ota}, and they will be the focus of this paper. 
One can easily expect that nothing stops us to apply the recursion relations in~\cite{Cheung:2015ota} for amplitudes containing particles other than scalars, as far as the amplitudes vanish fast enough in the soft limit. 
A typical such example will be considered in this paper is amplitudes of fermions in the Akulov-Volkov theory~\cite{Volkov:1972jx, Volkov:1973ix}, a effective field theory of goldstinos. 
Furthermore, we will extend the recursion relations for amplitudes with non-vanishing soft limits. 
In fact, the amplitudes with vanishing soft limits can be regarded as a special case of the non-vanishing limits. 
In our construction, it is not important whether the amplitudes vanish or not in the soft limit, what is essential is whether we have enough information of the amplitudes in the limit to overcome the bad large-$z$ behaviour.  
We will consider two general situations in this paper. 
One is that amplitudes contain only Goldstone bosons, but with non-vanishing soft limit. 
By performing ``soft-BCFW" shifts as done in~\cite{Cheung:2015ota}, and then applying Cauchy's theorem, one can obtain the amplitudes through a contour integral whose residues are all known, either from factorizations as the original BCFW recursion relations, or determined by the soft theorems.
The other more general case is that amplitudes contain not only Goldstone bosons but also other matter particles. 
For such general amplitudes, we perform mixed BCFW shifts on external momenta: soft-BCFW shifts on some Goldstone bosons and the original BCFW shifts on other legs~\footnote{The original BCFW shifts can be carried by Goldstone bosons or matter particles, for convenience and simplification of the calculation.}. 
We derive recursion relations again with the help of the soft theorems. 
For both cases, we study concrete examples as well. 
To illustrate the amplitude construction with pure Goldstone bosons and non-vanishing soft limits, we consider the dilaton effective action of the $a$-theorem~\cite{Komargodski:2011vj, Elvang:2012st}. 
For amplitudes with both Goldstone and non-Goldstone particles, we study the two-scalar model~\cite{Boels:2015pta} as an example. 
In both examples, the soft theorems are related to the conformal symmetry breaking, which have been studied recently in~\cite{Boels:2015pta, Huang:2015sla, DiVecchia:2015jaq}.

This paper is organized as following. 
We begin with section \ref{section:review} by briefly reviewing the recursion relations proposed in~\cite{Cheung:2015ota}.  
In section \ref{section:fermions} we apply the recursion relations of~\cite{Cheung:2015ota} for fermionic amplitudes in the Akulov-Volkov theory, and we compute the six-point amplitude using recursion relations as a concrete example. 
In section \ref{section:non-vanishingsoft}, we study amplitudes of pure Goldstone bosons, but with non-vanishing soft limits. 
Although zeros of amplitudes can not be arrived at in the soft limits, the soft theorems still provide enough information to build up recursion relations. 
We study amplitudes of dilaton effective action as an example to illustrate this idea. 
After that, we move on to section \ref{section:notjustdilaton} with a more general case where amplitudes contain not only Goldstone bosons but also other types of particles. 
A concrete example studied is the so-called two-scalar model. 
Finally, we close the paper with conclusions and remarks in section \ref{section:conclusion}.

\section{Scalar amplitudes with vanishing soft limits: a review} \label{section:review}

Amplitudes in effective field theories, especially those related to the symmetry breaking, typically have nice properties in the soft limits $p_i \rightarrow 0$. 
One famous example is the so-called ``Adler's zero", which states that amplitudes of Goldstone bosons from spontaneously global symmetry breaking vanish in the single soft limit~\cite{Adler:1964um}. 
This is precisely the important physical input utilized in~\cite{Cheung:2015ota} to obtain new recursion relations for amplitudes in effective field theories of pure scalars.  
This construction is realized by performing following all-line shifts, 
\bea \label{eq:all-line-shift}
p_{\hat{i}} = (1-a_i z) \, p_i \,.
\eea
Here $ \sum^n_{i=1} a_i \, p_i =0$, which is required by momentum conservation and can only have non-trivial solutions when $n>d+1$ (here $d$ is the dimension). 
We call such shifts as ``soft-BCFW" shifts. 
When the deformation parameter $z$ approaches $1/a_i$, the desired soft limit $\hat{p}_i \rightarrow 0$ is achieved. 
As original BCFW recursion relations, the amplitude $A_n(0)$ can be represented as a contour integral~\cite{Cheung:2015ota},
\bea \label{eq:contour-integral}
A_n(0) = {1 \over 2 \pi i} \oint_{|z|=0} {dz \over z} { A_n(z)  \over F_n^{(\sigma)}(z)} \,,
\eea
where the function $F_n^{(\sigma)}(z)$ is defined as
\bea
F_n^{(\sigma)}(z) = \prod^n_{i=1} (1-a_i z)^{\sigma} \, ,
\eea
with certain positive value $\sigma$. 
The purpose of introduced $F_n^{(\sigma)}(z)$ is to improve the large-$z$ behaviour of the constructed function ${ A_n(z)/F_n^{(\sigma)}(z)}$. 
In particular, if $\sigma$ is large enough, the constructed function definitely has a vanishing residue at $z=\infty$. 
On the other hand, $F_n^{(\sigma)}(z)$ might take a risk to introduce some additional poles (or even branch cuts for non-integer $\sigma$). 
Such construction can be useful only if we know the residues of the additional poles introduced by $F_n^{(\sigma)}(z)$. 
A special and simplest case was considered in~\cite{Cheung:2015ota}, where these additional poles have vanishing residues, in other words, the deformed amplitude $A_n(z)$ gives zeros at the poles $z={1 \over a_i}$, and those zeros precisely cancel the infinity from $F_n^{(\sigma)}(z)=0$ in the denominator. This is none other than amplitudes with vanishing soft limits. 

More concretely, if the amplitudes vanish in a certain power of the soft momentum as
\bea
A_n(\tau p_i) \sim \tau^{\sigma} \,, \quad {\rm with}  \quad \tau \rightarrow 0\,, 
\eea
then the shifted amplitude according to (\ref{eq:all-line-shift}) should have zeros at  $z={1 \over a_i}$, and goes as
\bea
A_n(z)\sim  (1- a_i z)^{\sigma} \, , \quad  {\rm with}  \quad z \rightarrow {1 \over a_i} \, ,
\eea
and these zeros precisely cancel the singularities from $F_n^{(\sigma)}(z)=0$. 
One can now apply the Cauchy's theorem to the contour integral (\ref{eq:contour-integral}), and obtain the recursion relations by summing over all simple poles from factorization diagrams of $A_n(z)$, 
\bea
A_n=A_n(0) = \sum_I {1 \over P_I^2} {A_L(z^-_I)A_R(z^-_I) \over (1-z^-_I/z^+_I) F_n^{(\sigma)}(z^{-}_I)} + (z^-_I \leftrightarrow z^+_I) \, ,
\eea
where $1/P^2_I$ is the internal propagator with $P_I = \sum_{i \in I} p_i$ and $z^{\pm}_I$ are the solutions to the massless on-shell condition
\bea
0=\big(P_I - z \sum_{i \in I} a_i p_i\big)^2 =P^2_I - 2z \big(\sum_{i \in I} a_i p_i\big) \cdot P_I +z^2 \big(\sum_{i \in I} a_i p_i\big)^2  \, .
\eea
The solution is given by
\bea \label{eq:solution}
z^{\pm}_I = {  \big(\sum_{i \in I} a_i p_i\big) \cdot P_I  \pm \sqrt{\big[ (\sum_{i \in I} a_i p_i) \cdot P_I\big]^2 - P^2_I \, \big(\sum_{i \in I} a_i p_i\big)^2 } \over \big(\sum_{i \in I} a_i p_i\big)^2 } \,. 
\eea
To really have the above recursion relations, an assumption that no pole appearing at $z=\infty$ is made, namely, if $A_n(z) \sim z^m$ at $z \rightarrow \infty$, one should have $m-n\sigma<0$ to have a vanishing residue at $z=\infty$ for the function ${ A_n(z)/F_n^{(\sigma)}(z)}$. 
Theories with enhanced soft limits~\cite{Cheung:2014dqa, Cheung:2015ota}, such as non-linear sigma model and scalar Dirac-Born-Infeld theory, exactly satisfy this criterion. 
Conceptually, the new recursion relations prove that amplitudes in a class of theories having bad large-$z$ behaviour are also on-shell constructible, with the help of enhanced soft limits. 
In practice, besides computing higher-point amplitudes, the recursion relations can provide a nice way to study new representations of scattering amplitudes, such as the CHY representation~\cite{Cachazo:2013hca}. 
For instance, they can offer a simple proof for amplitudes in the new representation by confirming their factorization properties, as well as their soft and large-$z$ behaviours. 
Finally, we remark that the soft-BCFW shifts (\ref{eq:all-line-shift}) can also apply to amplitudes with multiple soft legs. 
This can be realized by setting some of the parameters $a_i$ in (\ref{eq:all-line-shift}) to be equal, such that more than one momentum simultaneously approach to zero in the limit $z \rightarrow 1/a_i$. 

From the above discussion it is clear that although the recursion relations were originally only applied to scalar amplitudes in~\cite{Cheung:2015ota}, such restriction is not necessary as far as the amplitudes have the enhanced soft behaviour. Indeed there exist amplitudes of particles other than scalars which also have the enhanced soft limits. One particular example~\cite{Chen:2014xoa} is the fermionic amplitudes in the Akulov-Volkov theory~\cite{Volkov:1972jx, Volkov:1973ix}, namely the effective theory of spontaneously supersymmetry breaking which describes the interaction of the Goldstinos\footnote{The Akulove-Volkov action gives a nonlinear realization of supersymmetry. One the other hand, the low-energy effective Lagragian of Goldstinos can be constructed by a nilpotent constraint of the chiral superfield~\cite{Komargodski:2009rz}. One can prove that the Lagrangian from constraint can be transferred to the Akulove-Volkov Lagrangian after a field redefinition~\cite{Luo:2009ib, Luo:2009pz,Luo:2010zp,Liu:2010sk, Liu:2011ni}. }. We will study the amplitudes in the Akulove-Volkov theory in the next section. 
More importantly, having vanishing soft limits is actually also not a necessary requirement.
In fact, we will consider an example in which the amplitudes are even singular in the soft limits. 
The recursion relations can be established as far as we have enough knowledge of the amplitudes in the soft limits, in other words, the residues at the additional poles introduced by the function $F_n^{(\sigma)}(z)$ can be obtained from soft theorems. 
Those theories with vanishing soft limits are special since the residues of the poles at $F_n^{(\sigma)}(z)=0$ vanish. 
These more interesting generalizations with non-vanishing soft limits will be considered in the following sections after the discussion the on-shell construction of fermionic amplitudes with the enhanced soft limits, {\it i.e.} amplitudes in the Akulov-Volkov theory. 

\section{Akulov-Volkov amplitudes from recursion relations} \label{section:fermions}

In this section, we apply the recursion relations reviewed in the previous section to amplitudes of effective field theories containing fermions, in particular, the Akulov-Volkov theory~\cite{Volkov:1972jx, Volkov:1973ix}. 
The Akulov-Volkov theory is an effective theory of $\mathcal {N} =1$ supersymmtry spontaneously breaking.
The action can be written as 
\bea
S_{\rm AV} = -{1 \over 2 g^2} \int d^4x \,  {\rm det} \,  (1 + i g^2 \psi \sigma^{\mu} \overleftrightarrow{\partial}_{\mu} \bar{\psi} ) \, ,
\eea
where $\psi$ (and $\bar{\psi}$) is the Goldstino. 
Non-vanishing amplitudes contain the same numbers of $\psi$ and $\bar{\psi}$, i.e., $A( \bar{\psi}_1, \psi_2, \ldots, \bar{\psi}_{n-1}, \psi_{n}  )$. 
Unlike the scalars, fermions have non-trivial wave functions, therefore the amplitudes behave differently according to how the momentum is taken to be soft. 
If the soft limit $p_i \rightarrow \tau p_i$ is realized via
\bea \label{eq:fermion-shift-1}
\lambda_{i} \rightarrow \tau \lambda_{i}, \quad  {\rm if} ~i=2k-1; \quad {\rm or} \quad 
\tilde{\lambda}_{i} \rightarrow \tau \tilde{\lambda}_{i},\quad  {\rm if} ~i=2k \, ,
\eea
then wave functions in the amplitude $A( \bar{\psi}_1, \psi_2, \ldots, \bar{\psi}_{n-1}, \psi_{n}  )$ do not carry a factor of $\tau$, since the wave function $\psi \sim \lambda$ and $\bar{\psi} \sim \tilde{\lambda}$.
Here we use standard notation of spinor-helicity formalism (for a comprehensive review, please see~\cite{Elvang:2013cua}), where the null momentum can be expressed as a product of two spinors, 
\bea
p^{\mu}_i \sigma_{\mu}^{\alpha, \dot{\alpha}} = \lambda^{ \alpha }_i \tilde{ \lambda}^{ \dot{\alpha} }_i \, ,
\eea
and the Lorentz invariants can be built from $ \lambda^{ \alpha }$ and $\tilde{ \lambda}^{ \dot{\alpha} }$, 
\bea
\langle i \, j \rangle := \epsilon_{\alpha \, \beta} \lambda^{ \alpha }_i \lambda^{ \beta }_j \, , \quad 
[ i \, j ] := \epsilon_{ \dot{\alpha} \, \dot{\beta}}
\tilde{ \lambda}^{ \dot{\alpha} }_i \tilde{ \lambda}^{ \dot{\beta} }_j \, ,
\eea
and the Mandelstam variables $s_{ij}=(p_i+p_j)^2$ can be written as $s_{ij} = \langle i\,j\rangle [i\,j]$.
With the ``soft-BCFW" shifts given in (\ref{eq:fermion-shift-1}), in the single-soft limit the amplitude behaves as~\cite{Chen:2014xoa}
\bea
A_n (\tau \lambda_{{2k-1}} ~ {\rm or} ~ \tau \tilde{\lambda}_{2k}) \sim  \tau^1 \,  \quad {\rm with} \quad \tau \rightarrow 0 \, .
\eea
Corresponding to this type of soft rescaling, we take the following BCFW shifts, 
\bea \label{eq:fermion-BCFW}
\lambda_{\widehat{2k-1}} = \lambda_{2k-1} ( 1 - a_{2k-1} z ) \, , \quad
\tilde{\lambda}_{\widehat{2k}} = \tilde{\lambda}_{2k} ( 1 - a_{2k} z ) \, .
\eea
Analogous to the pure-scalar case, the recursion relations can be obtained through the contour integral representation of the amplitude, 
\bea \label{eq:contour}
A_n(0)={1 \over 2 \pi i}\oint_{|z|=0} { A_n (z) \over z F^{(1)}_n(z) } \, .
\eea
As $A_n (z) \sim z^{n/2}$ in the large-$z$ limit (which can be seen by a simple dimension power counting), the above constructed contour integral has no residue at $z=\infty$ by introducing $F^{(1)}_n(z)=\prod_{i=1}^n (1-a_i z)$ in the denominator. Through Cauchy's theorem, we derive the recursion relations for the amplitudes in the Akulov-Volkov theory, which hold the same form as those of the scalar amplitudes,  
\bea
A_n = \sum_I {1 \over P_I^2} { A_L (z_I^-)A_R (z_I^-) \over (1 - z_I^-/z_I^+ )F^{(1)}_n(z_I^-) } + ( z^-_I \leftrightarrow z^+_I ) \, .
\eea
One comment before we move on: one can of course choose any other different shifts to achieve the soft limits, and they should work as well. 
For instance, one may shift $\lambda$'s only as, 
\bea \label{eq:fermion-BCFW2}
\lambda_{\widehat{i}} = \lambda_{i} ( 1 - a_{i} z ) \,  .
\eea
With such shifts, the soft behaviour of amplitudes changes because wave functions corresponding to ${\psi}$'s now get an extra factor of $( 1 - a_{i} z ):=\tau$,
\bea
A_n(\tau p_{2k-1}) \sim  \tau^2 \, , \quad A_n(\tau p_{2k}) \sim  \tau^1 \, , \quad {\rm with} \quad \tau \rightarrow 0 \, .
\eea
Correspondingly, the contour integral should be modified as well,
\bea \label{eq:contour2}
A_n(0)={1 \over 2 \pi i}\oint_{|z|=0} { A_n (z) \over z F^{(\sigma_i)}_n(z) } \, ,
\eea
where $F^{(\sigma_i)}_n(z)=\prod^n_{i=1} ( 1- a_i z)^{\sigma_i }$, with $\sigma_i=1$ if $i$ is even and $\sigma_i=2$ if $i$ is odd. 

As a concrete example, we finish this section by investigating a six-point amplitude in this theory from the recursion relations. 
Using the shifts of (\ref{eq:fermion-BCFW}), we gain the six-point amplitude from the recursion relations
\bea\label{VA-6pt}
A_6 = \sum_I {1 \over P_I^2} { A_L (z_I^-)A_R (z_I^-) \over (1 - z_I^-/z_I^+ )F^{(1)}_6(z_I^-) } + ( z^-_I \leftrightarrow z^+_I ) \, ,
\eea
where $A_L(z)$ and $A_R(z)$ are four-point amplitudes with shifted momenta. 
The four-point amplitude is given as\footnote{Here and in the following we set $g^2=1/2$ such that $g$ will not be carried in all formulas.}
\bea
A_4( \bar{\psi}_1, \psi_2, \bar{\psi}_3 , \psi_{4}) =  s_{13} \langle 2\,4\rangle [1\,3] \, .
\eea
Note that the four-point amplitude is in fact completely fixed by the mass dimension, the little group scaling and the (anti-)symmetry of exchanging $1 \leftrightarrow 3$ and $2 \leftrightarrow 4$. 
Let us focus on a particular factorization channel, e.g., ${1/s_{123}}$. 
Substitute the four-point amplitudes into (\ref{VA-6pt}), we find the expression for this channel is given by
\bea \label{eq:channel123}
A^{(123)}_6 = {[1\,3]\langle 46\rangle  \langle 2 |p_1(z^-_{123})+p_3(z^-_{123})| 5]  \over s_{123}} { s_{13} (z_{123}^-) s_{46} (z_{123}^-) \over (1 - z_{123}^-/z_{123}^+ )F^1_6(z_{123}^-) } + ( z^-_{123} \leftrightarrow z^+_{123} ) \, ,
\eea
with the solution of $z^{\pm}_{123}$ given in (\ref{eq:solution}), now with $i \in \{1,2,3\}$. 
One useful trick to simplify the calculation is to rewrite $A^{(123)}_6$ as a residue, and then apply the Cauchy's theorem. The result in (\ref{eq:channel123}) can be re-expressed as
\bea
A^{(123)}_6 =- [1\,3]\langle 46\rangle  
\, {\rm Re}_{z^{\pm}_{123}} \left[\langle 2 |p_1(z)+p_3(z)| 5]  { s_{13}(z) s_{46}(z) \over z s_{123}(z) F^{(1)}_6(z) } \right] \, ,
\eea
applying the residue theorem, we have
\bea
A^{(123)}_6=
[1\,3]\langle 46\rangle  \langle 2 |1+3| 5] 
{s_{13}s_{46} \over s_{123}} -\sum_{i} R^{(123)}_i \, ,
\eea
where the first term comes from the residue at the pole $z=0$, and the remaining terms $\sum_{i}R^{(123)}_i$ are the sum of the residues at poles from $F^{(1)}_6(z)=0$ of $z=1/a_i$. 
It turns out that after summing over all the factorization channels, the residues from $F^{(1)}_6(z)=0$ all nicely cancel out. 
First of all, it is easy to see that for the channel of $s_{123}$ only $R^{(123)}_2$ and $R^{(123)}_5$ are non-vanishing, due to the factor $ s_{13}(z) s_{46}(z)$ in the numerator. We observe the following cancellation, 
\bea
R^{(123)}_2 + R^{(125)}_2 + R^{(235)}_2 =0 \, , \cr
R^{(123)}_5 + R^{(134)}_5 + R^{(136)}_5 =0 \,, 
\eea
namely the residue of the pole $z=1/a_2$ for channel $1/s_{123}$ is cancelled by that from the channels $1/s_{125}$ and $1/s_{235}$ at the same pole, and the same for the residue of $z=1/a_5$. So eventually, the full amplitude can be expressed in a very compact form, 
\bea
A_6 = [1\,3]\langle 46\rangle  \langle 2 |1+3| 5] {s_{13}s_{46} \over s_{123}}
+ \ldots \, ,
\eea
where $\ldots$ indicates summing over all other channels except the one with propagator $1/s_{135}$ since all the $\psi$ cannot on one side of the diagram. Due to the permutation symmetry among $\{\psi_1, \psi_3, \psi_5 \}$ as well as  among $\{\bar{\psi}_2, \bar{\psi}_4, \bar{\psi}_6 \}$, the contributions from all other channels can be obtained from the result of the channel $1/s_{123}$ by simply exchanging indices. For instance, the contribution of the channel $1/s_{124}$ can be obtained from that of $1/s_{123}$ with $1 \leftrightarrow 5, 2 \leftrightarrow 6$, which leads to
\bea
[3\,5]\langle 24\rangle  \langle 6 |3+5| 1] 
{s_{35}s_{24} \over s_{124}} \, .
\eea
We have checked the agreement of this result with explicit Feynman diagram calculation~\cite{Chen:2014xoa}.

In this section we generalize the recursion relations in~\cite{Cheung:2015ota} to amplitudes containing fermions. 
Clearly, one can make further generalization to amplitudes with other particle species, as far as the amplitudes behave nicely in the soft limits which can help to overcome the bad large-$z$ behaviour. 
It would be even more interesting to consider possible generalization of the recursion relations with supersymmetry, quite similar to the generalization of the original BCFW recursion relations to supersymmetric theories~\cite{Luo:2005rx, Brandhuber:2008pf, ArkaniHamed:2008gz}. 
For the supersymmetrization of the recursion relations, it would require a better understanding on soft limits of super amplitudes, which we will leave as a future research direction. 
Instead, in the following sections we will study amplitudes with non-vanishing soft limits.

\section{Amplitudes with non-vanishing soft limits}  \label{section:non-vanishingsoft}
In this section, we consider the amplitudes which do not vanish in the single soft limit, but still have an universal behaviour \bea \label{eq:general-soft-theorem}
A_n(\tau p_n){\big |}_{\tau \rightarrow 0} = 
 \sum^{q_2}_{k=q_1} \tau^k \left( \mathcal{S}^{(k)}_n A_{n-1} \right) + \mathcal{O}(\tau^{q_2+1}) \, .
\eea 
The soft limit of an $n$-point amplitude is universally given by soft factors $\mathcal{S}^{(k)}_n$ acting on an $(n{-}1)$-point amplitude\footnote{Here we assume the soft theorem is formed as soft factors acting on a lower-point amplitude, although this assumption is not necessarily required for the following derivation of the recursion relations.}, which starts from the order $\mathcal{O}(\tau^{q_1})$ and contains known information up to the order $\mathcal{O}(\tau^{q_2})$. 
Here the subscript $n$ in $\mathcal{S}^{(k)}_n$ denotes that particle $n$ is taken to be soft.
We perform the same soft-BCFW shifts on all external momenta,
\bea
p_{\hat{i}} = p_i (1 - a_i z) \, ,
\eea
and the new recursion relations will be derived by considering the un-shifted $A_n(0)$ as the following contour integral as before, 
\bea \label{eq:general-contour-integral}
A_n(0) = {1 \over 2 \pi i} \oint_{|z|=0} {dz \over z} { A_n(z)  \over F^{( \sigma )}_n(z)} \,,
\eea
with $\sigma=q_2+1$. 
The $\sigma$ is chosen such that all the residues corresponding to the poles from $F_n^{( \sigma )}(z)$ can be extracted from the soft theorem (\ref{eq:general-soft-theorem}). 
To ensure the recursion relations establishable, the large-$z$ behaviour of the amplitude $A_n(z) \sim z^m$ is required to satisfy the following condition, 
\bea
m < n\,(q_2+1) \,.
\eea
This is an on-shell constructible criterion for the amplitudes accompanying with soft theorems.
Applying the Cauchy's theorem to (\ref{eq:general-contour-integral}), we obtain the recursion relations with the help of the soft theorem (\ref{eq:general-soft-theorem}),
\bea\label{eq:recursion_general} 
A_n(0) = \sum_I{}^{'} {1 \over P_I^2} { A_L (z_I^-)A_R (z_I^-) \over (1 - z_I^-/z_I^+ )F_n^{(\sigma)}(z_I^-) } + ( z^-_I \leftrightarrow z^+_I ) - \sum^n_{i=1} \sum^{q_2}_{k=q_1} R^{(k)}_i \, .
\eea
The first term with the ``primed sum" corresponds to factorization poles, but excluding channels containing a three-point amplitude, the reason will become clear shortly. 
The terms $R^{(k)}_i$'s are non-trivial contributions from the poles at $z = {1 \over a_i}$, which can be completely determined by the soft theorem (\ref{eq:general-soft-theorem}), 
\bea
R^{(k)}_i = {1 \over 2 \pi i}\oint_{z={1 \over a_i}} dz {\left(\mathcal{S}^{(k)}_i A_{n-1, i}\right) (z) \over z (1- a_i z)^{q_2+1-k} F_{n-1,i}^{(q_2+1)}(z)} \, ,
\eea
where $A_{n-1, i}$ is an $(n{-}1)$-point amplitude with the leg $i$ removed, and 
\bea
F_{n-1, i}^{(q_2+1)}(z) = { F_{n}^{(q_2+1)}(z) \over (z-a_i)^{(q_2+1)} } \, .
\eea
Furthermore $\left(\mathcal{S}^{(k)}_i A_{n-1, i}\right)(z)$ means that the soft-BCFW shifts are imposed on the results obtained from the soft operators $\mathcal{S}^{(k)}_i$ acting on the lower point amplitude $A_{n-1, i}$.  
The above expression is obtained by using the soft theorems (\ref{eq:general-soft-theorem}) since $p_{\hat{i}} \rightarrow 0$ when $z \rightarrow {1 \over a_i}$. 
For the cases $q_2+1-k > 1$, the pole $1- a_i z=0$ is of a higher order, so to obtain the residue one needs to expand $\left(\mathcal{S}^{(k)}_i A_{n-1, i}\right) (z)$ to order $(z-1/a_i)^{q_2-k}$. 
Finally, as we mentioned that in the first term of the recursion relations (\ref{eq:recursion_general}) we do not sum over the factorization channels containing a three-point amplitude because they have been included in the soft-theorem contributions $R^{(k)}_i$. 
Concretely saying, the poles of such factorization channels are, 
\bea
{1 \over s_{\hat{i} \hat{j} }} = {1 \over s_{ij} (1- a_i z)(1 - a_j z)} \, ,
\eea
thus these three-point amplitudes only contain the poles of the form $z = {1 \over a_i}$ under the soft-BCFW shifts and have been included in $R^{(k)}_i$. 

In the following discussion, we apply the above general formulation to the amplitudes from the dilaton effective actions, which were first derived in~\cite{Komargodski:2011vj} in 4 dimension to prove the $a$-theorem and later generalized in~\cite{Elvang:2012st, Elvang:2012yc} to higher dimensions. 
The effective actions are constructed through the most general diffeomorphism invariant actions yielding the trace anomaly under the Weyl transformation, where the dilaton shifts as $\tau\rightarrow \tau+\sigma$. 
Generally, the trace anomaly obeys  
\bea
\langle T^\mu\,_\mu\rangle=\sum_{i}c_iI_i-(-)^{\frac{d}{2}}a\,E_d
\eea
where $E_d$ is the $d$-dimensional Euler density and $\sqrt{-g}I_i$ are conformal invariants. 
The coefficient $a$ serves as a candidate that monotonously decreases along a RG flow from UV to IR, and that is the content of the $a$-theorem. 
In the IR, the massless degrees of freedom include a dilaton $\tau$ which is the Goldstone boson from the spontaneously conformal symmetry breaking (or a compensating field if the conformal symmetry is explicitly broken). 
The dynamics of the dilaton is then governed by the dilaton effective action, which is defined as a diffeomorphism invariant functional $\Gamma[g,\tau]$, such that under the Weyl transformation it generates the required anomaly,
\bea
\delta \Gamma[g,\tau]_{\rm dilaton}=\int d^d x\sqrt{-g}\left(\sum_{i}c_iI_i-(-)^{d \over 2} \Delta a\, E_d\right)\,,
\eea
where $\Delta a = a_{\rm uv} - a_{\rm IR}$. The explicit form of the dilaton effective action is given in~\cite{Elvang:2012yc}.

Some explicit amplitudes have been computed in~\cite{Elvang:2012st} for $d=6$, here we list some of them, 
\bea
\nonumber 
A_4^{(4)}=\frac{b}{2f^8}(s_{12}^2+s_{13}^2+ s_{14}^2),&&\quad A_4^{(6)}=\left[\frac{3\Delta a}{2}-\frac{b^2}{f^4}\right]\frac{3}{f^8}s_{12}\, s_{13}\, s_{14} \, , \\
A^{(4)}_5=\frac{3b}{4f^{10}}\sum_{1\leq i<j\leq5}s^2_{ij},&&\quad A_5^{(6)}=\left[\frac{3\Delta a}{2}-\frac{b^2}{f^4}\right]\frac{2}{f^{10}}\sum_{1\leq i<j\leq5}s^3_{ij} \, ,
\eea
where the superscript of a $n$-point amplitude $A^{(m)}_n$ indicates the mass dimension of the amplitude, with $m \leq n+2$. 
More higher-point amplitudes have been computed \cite{Huang:2015sla} and can be found in the Appendix A of that paper. 
The soft theorems of the amplitudes in the dilaton effective action were first studied in~\cite{Huang:2015sla}, i.e.,  
\bea \label{eq:dilaton}
A_n(\tau p_n){\big |}_{\tau \rightarrow 0} = 
 \mathcal{S}^{(0)}_n A_{n-1} (p_1, \ldots, p_{n-1})  + \mathcal{O}(\tau^{1}) \, ,
\eea 
where the subscript in the soft factor $\mathcal{S}^{(0)}_n$ indicates which leg is taken to be soft (here, leg $n$ has been taken as soft), and $\mathcal{S}^{(0)}_n$ is given by
\bea\label{eq:softnomassS0}
\mathcal{S}^{(0)}_n = \sum^{n-1}_{i=1}
 \left( p_i \cdot {\partial \over \partial p_i } + {d- 2 \over 2} \right) - d \, .
\eea
The soft theorems were derived from the Ward identity of conserved currents, and further confirmed by concrete examples up to eight-point amplitudes~\cite{Huang:2015sla}. 
To derive the recursion relations, we again construct the amplitudes as a contour integral, 
\bea
A^{(m)}_n(0) = {1 \over 2 \pi i}\oint_{|z|=0} dz {A^{(m)}_n(z) \over z F^{ (\sigma) }_n(z)} \, .
\eea
Clearly the amplitude $A^{(m)}_n$ under the soft-BCFW shifts (\ref{eq:all-line-shift}) goes as $z^m$ in the limit $z\rightarrow \infty$. 
So if $m \geq n$, $\sigma>1$ is required, and the leading order soft theorems in (\ref{eq:dilaton}) cannot provide enough information for these cases. 
Thanks to very recent results in~\cite{DiVecchia:2015jaq}, general soft theorems for dilaton from spontaneously conformal symmetry breaking were derived, which go beyond the leading order $\mathcal{S}^{(0)}$. 
The universal soft theorems to the subleading order take the form, 
\bea \label{eq:softlimitCDBI}
A_n{\big |}_{p_n \rightarrow \tau p_n} \rightarrow  
\left(  \mathcal{S}^{(0)}_n + \tau \mathcal{S}^{(1)}_n \right) A_{n-1}(p_1 , \ldots, \overline{p}_{n-1}) + \mathcal{O}(\tau^2) \, ,
\eea
where the subleading soft operator $\mathcal{S}^{(1)}_n$ is given as~\cite{DiVecchia:2015jaq}
\bea\label{eq:softnomassS1}
\mathcal{S}^{(1)}_n  =   p^{\mu}_n \sum^{n-1}_{i=1} 
\left[ {1 \over 2} \left( 2 \, p^{\nu}_i { \partial^2 \over \partial {p_i^{\nu}} \partial {p_i^{\mu}} }
-
{p_{i\mu}}{ \partial^2 \over \partial {{p_i}_{\nu}} \partial {p_i^{\nu}} } \right) 
+ {d-2 \over 2} {\partial \over \partial {p^{\mu}_i} } \right] \, .
\eea
Furthermore one should remove one of the momenta in $A_{n-1}$ by the momenta conservation, that is because $\mathcal{S}_n^{(1)}$ is a non-trivial differential operator acting on $p_i$'s, so one should use independent variables. In (\ref{eq:softlimitCDBI}), we remove $p_{n-1}$, and replace it by $\overline{p}_{n-1} = - (\sum^{n-2}_{i=1} p_i)$. 
We have explicitly checked the soft theorems against the amplitudes computed in~\cite{Huang:2015sla}.  
With the results of the subleading soft factor on hand, we are able to derive the recursion relations from the following contour integral, 
\bea
A_n(0) = \oint_{|z|=0} dz {A_n(z) \over z F^{(2)}_n(z)} \, .
\eea
So this is just a special case of what we have discussed in the general formulation (\ref{eq:general-contour-integral}) with $q_1=0$ and $q_2=1$. 
Thus the recursion relations are given by
\bea \label{eq:recursiondilaton}
A_n = \sum_I{}^{'} {1 \over P_I^2} { A_L (z_I^-)A_R (z_I^-) \over (1 - z_I^-/z_I^+ )F_n^{(2)}(z_I^-) } + ( z^-_I \leftrightarrow z^+_I ) - \sum^n_{i=1} \sum^1_{k=0} R^{(k)}_i \, ,
\eea
with $R^{(k)}_i$ as
\bea \label{eq:Ri2}
R^{(0)}_i = {1 \over 2 \pi i}\oint_{z = 1/a_i} dz {\left(\mathcal{S}^{(0)}_i A_{n-1,i}\right)(z) \over z F^{(2)}_n(z) } \, , \quad 
R^{(1)}_i ={1 \over 2 \pi i} \oint_{z = 1/a_i}  dz {\left(\mathcal{S}^{(1)}_i A_{n-1,i}\right)(z)  \over  z (1-z a_i) F^{(2)}_{n-1, i}(z) } \, .
\eea
As before, $F^{(2)}_{n-1, i}(z)$ is simply $F^{(2)}_{n}(z)$ with the term $(1- a_i z)^2$ removed, namely $F^{(2)}_{n-1, i}(z) = \prod^n_{j=1, j\neq i} (1- a_j z)^2$, and $A_{n-1,i}(z)$ denotes the $(n{-}1)$-point amplitude with shifted momenta $( {p}_{\hat1}, \ldots, {p}_{\widehat{i-1}}, {p}_{\widehat{i+1}}, \ldots, {p}_{\hat n})$. 
The integrand of the first term $R^{(0)}_i$ contains double pole at $z=1/a_i$, so we need to expand the known function $\left(\mathcal{S}^{(0)}_i A_{n-1,i}\right)(z)$ to order $\mathcal{O}(1- a_i z)$, while for the second term $R^{(1)}_i$, one can simply plug $z=1/a_i$ in $\left(\mathcal{S}^{(1)}_i A_{n-1,i}\right)(z)$. 


Here we compute analytically five-point amplitude from four-point ones. 
Due to the momenta conservation constraint 
\bea \label{eq:constraint-on-ai}
\sum^{n}_{i=1} a_i \, p_i =0 \, ,
\eea
those $a_i$ can only have non-trivial solutions in $n > d+1$ for dimension $d$. 
For this special five-point amplitude, we will consider the amplitude in the three-dimensional kinematics. 
Since any factorization diagram of a five-point amplitude will necessarily contain a three-point amplitude, which has been included in the soft contributions as we have discussed. Thus the recursion relations only gain contributions from (\ref{eq:Ri2}), in other words, from the soft operators acting on the four-point amplitude. 
For $n=5$, (\ref{eq:constraint-on-ai}) can be solved by following $a_i$ $(i=1,\cdots,5)$
\bea\label{DBIA5p6ai}
&&a_1=s_{23} (s_{14} s_{23} - s_{13} s_{24} - s_{12} s_{34})\,, \quad
a_2=s_{13} (-s_{14} s_{23} + s_{13} s_{24} - s_{12} s_{34})\, , \cr
&&a_3=s_{12} (-s_{14} s_{23} - s_{13} s_{24} + s_{12} s_{34})\, , \quad
a_4=2s_{12} s_{13} s_{23}\,, \quad 
a_5=0 \, .
\eea
To obtain the results of $R_i^{(k)}$, we start from the four-point amplitude expression
\bea
A_4^{(6)}(1,2,3,4)= \left[{3\Delta a\over 2}-{b^2\over f^4}\right] {3\over f^8} s_{12}\,s_{13}\,s_{14},
\eea
apply the soft operators on it, and then perform the soft-BCFW shifts.
For instance, for $R_1^{(k)}$, we find,
\bea
R^{(0)}_1 = {1 \over 2 \pi i}\oint_{z = 1/a_1} dz {\left(\mathcal{S}^{(0)}_1 A_{4,1}^{(6)}\right)(z) \over z F^{(2)}_5 (z) }=-\left[{3\Delta a\over 2}-{b^2\over f^4}\right] {12\over f^8}{(1-a_5/a_1)\over(1-a_5/a_1)^3}\,s_{23}\,s_{34}\,s_{24}
\eea
and
\bea
R^{(1)}_1 &=&{1 \over 2 \pi i} \oint_{z = 1/a_1} dz {\left(\mathcal{S}^{(1)}_1 A_{4,1}^{(6)}\right)(z) \over z F^{(2)}_5 (z) } 
\cr
&=&
-{1\over (1-a_5/a_1)^2}\left[{3\Delta a\over 2}-{b^2\over f^4}\right] {3\over 2 f^8}\cr
&\times&\bigg\{\,s_{12} s_{34} \bigg[3{s_{23}\over (1-a_4/a_1)}+3{ s_{24}\over (1-a_3/a_1)} -{s_{34}\over (1-a_2/a_1)}\bigg]
+ 
(2 \leftrightarrow 3) + ( 2 \leftrightarrow 4) \,\bigg\} \, .
\eea
Due to the specific choice of $a_i$ in (\ref{DBIA5p6ai}), the factor $(1-a_5/a_i)$ simply goes to 1 since $a_5=0$. 
Moreover, there are no contributions from $R_5^{(0)}$ and $R_5^{(1)}$. 
By summing over all $R^{(0)}_i$ and $R^{(1)}_i$ ($i=1,\cdots,4$), we obtain the five-point amplitude
\bea
A_5^{(6)} =-\sum_{i=1}^4 \left(R_i^{(0)}+R_i^{(1)}\right)= \left[{3\Delta a\over 2}-{b^2\over f^4}\right]{2\over f^{10}} \sum_{1\leq i<j\leq 5} s_{ij}^3 \,.
\eea 
Interestingly, the above result is even valid in $d>3$ dimension. It would be very interesting to investigate further how general it is that a computation in a lower dimension can be directly promoted as a result in higher dimensions. In fact if we change the form of the four-point amplitude or choose a different solution for $a_i$ (in particular those with $a_5 \neq 0$), we find the result will be only strictly valid for three-dimensional kinematics. We have further tested the recursion relations numerically for amplitudes up to seven points.

We finally comment that the same recursion relations (\ref{eq:recursiondilaton}) can apply to the action of a single D$_{(d-1)}$-brane in the AdS$_{d+1}$ space, which is described by the conformal DBI action
\bea
S_{\rm CDBI}  = -\int d^d x \, {1 \over \phi^d } \left( \sqrt{  1 + {\partial_{\mu} \phi \partial^{\mu} \phi  }}  -1\right) \, ,
\eea
where the scalar $\phi$ corresponds to the radius coordinate of AdS space, and its relation to the dilaton can be found in~\cite{Elvang:2012st}. 
The action is related to the effective action of the dual CFT ($\mathcal{N}=4$ super Yang-Mills theory when $d=4$) in the Coulomb branch. The conformal symmetry of the action is spontaneously broken by giving the scalar a vacuum expectation value (vev), $\phi=v+\delta \phi$. The action should be understood as polynomials of $\delta \phi$ and $\partial_{\mu} \delta \phi$ by expanding $1/\phi^d$ and the square root term. Some of the matrix elements of this action have been computed in~\cite{Elvang:2012st}. In fact, amplitudes from the conformal DBI action are a special case of those from dilaton effective action, with $\Delta a = {2 \over 3} {b^2 \over f^4}$. 

It is straightforward to see that with the condition $\Delta a = {2 \over 3} {b^2 \over f^4}$, amplitudes $A^{(m)}_n$ in conformal DBI with $m > n$ actually vanish. For the special case of $m=n$, the amplitudes are the same as those from the usual flat space DBI action. 
It can be obtained by simply replacing the factor $\phi^d$ with the vev $v^d$. 
We can also understand that the enhanced soft behaviour of the flat space DBI action as emphasized in~\cite{Cheung:2014dqa, Cheung:2015ota} is a special case of the general soft theorems (\ref{eq:softlimitCDBI}), due to the fact that the amplitude $A^{(n)}_{n-1}(z)$ vanishes for the flat space DBI action, and we have
\bea \label{eq:softlimitDBI}
A^{(n)}_n{\big |}_{p_n \rightarrow \tau p_n} = 
\left( \mathcal{S}^{(0)}_n + \tau \mathcal{S}^{(1)}_n \right) A^{(n)}_{n-1} + \mathcal{O}(\tau^2) = \mathcal{O}(\tau^2) \, .
\eea
In this sense, the universality of the soft theorems (\ref{eq:softlimitCDBI}) to the subleading order is closely related to the enhanced soft behaviour of the flat space DBI action. 

\section{Amplitudes with non-Goldstone bosons} \label{section:notjustdilaton}

A more general case is that amplitudes contain other particles rather than only Goldstone bosons, and the soft theorems by taking a single Goldstone boson to be soft are valid for such amplitudes as well. 
In this section, we propose recursion relations for such general situation. 
Apparently the all-line shift in (\ref{eq:all-line-shift}) is not suitable any more since generally non-Goldstone particles fail to provide any useful information in the soft limit $z \rightarrow 1/a_i$. 
Instead, we perform the soft-BCFW shifts (\ref{eq:all-line-shift}) only on Goldstone bosons, and do the usual BCFW shifts (\ref{eq:usualBCFW}) on other legs which can be any particles including Goldstone bosons. 

To illustrate this idea, we take the two-scalar model as an example. 
This model was considered in~\cite{Boels:2015pta} to study the soft theorems from spontaneously conformal symmetry breaking. 
The Lagrangian is given as
\bea
\mathcal {L} = {1 \over 2} \partial_{\mu} \xi \partial^{\mu} \xi  
+
{1 \over 2} \partial_{\mu} \phi \partial^{\mu} \phi
-{1 \over 2} \lambda^{4 \over d-2} \phi^2 (\xi + v)^{ 4 \over d-2} \, ,
\eea
where $v$ is the vev of the dilaton field $\xi$, and one can expand $(\xi + v)^{ 4 \over d-2}$ to obtain the interactions, which could have infinity terms. The mass of the scalar $\phi$ is given by 
\bea
m^2 =  (\lambda \, v)^{ 4 \over d-2} \, .
\eea
For the simplicity in the following we will set the vev $v=1$. 
We now consider amplitudes with at least one dilaton, at the position $1$, denoted as $A_n\,(\xi_1, \ldots)$, and ``$\ldots$" means other particles can be any possible species, namely $\phi$ or $\xi$. 
The amplitudes in this model satisfy the soft theorems discussed in section \ref{section:non-vanishingsoft}, but including more soft factors to deal with the massive particles~\cite{Boels:2015pta, DiVecchia:2015jaq}, 
\bea \label{eq:softdouble-scalar}
A_n{\big |}_{p_1 \rightarrow \tau p_1} \rightarrow  
\left( \tau^{-1} \mathcal{S}^{(-1)}_{{\rm M}, 1} + \mathcal{S}^{(0)}_{{\rm M}, 1}  + \tau \mathcal{S}^{(1)}_{{\rm M}, 1}
+ \mathcal{S}^{(0)}_1 + \tau \mathcal{S}^{(1)}_1 \right) A_{n-1, 1}(p_2 , \ldots, \bar{p}_{n}) + \mathcal{O}(\tau^2) \, ,
\eea
where $p_1$ is taken to be soft. 
The soft operators $\mathcal{S}^{(0)}_1$ and $\mathcal{S}^{(1)}_1$ are independent of mass, and are the same as (\ref{eq:softnomassS0}) and (\ref{eq:softnomassS1}) by switching the soft particle from leg $n$ to leg $1$. 
The soft factors $\mathcal{S}^{(i)}_{{\rm M}, 1}$ proportional to the mass $m_i$ of each external leg are given as
\bea
\mathcal{S}^{(-1)}_{{\rm M}, 1} &=& \sum^{n}_{i=2} {m_i^2 \over p_1 \cdot p_i}  \, , 
\quad \mathcal{S}^{(0)}_{{\rm M}, 1} = \sum^{n}_{i=2} {m_i^2 \over p_1 \cdot p_i} p_1 \cdot { \partial \over \partial p_{i} } \, , \cr
\mathcal{S}^{(1)}_{{\rm M}, 1} &=& {1 \over 2} 
\sum^{n}_{i=2} {m_i^2 \over p_1 \cdot p_i} p^{\mu}_1 p_1^{\nu} \, 
{ \partial^2 \over \partial p^{\mu}_{i}  \partial p^{\nu}_{i}} \, .
\eea
As there is no derivative interaction in this two-scalar model, the amplitudes go worst as $z^0$ in the large-$z$ limit. 
So the bad large-$z$ behaviour can be removed by dividing a simple factor such as $F^{(1)}_1(z)=(1-z)$. 
Instead of all-line soft-BCFW shifts preformed in section \ref{section:non-vanishingsoft}, here we consider shifts as 
\bea
p_{\hat{1}} = (1 - z) p_1 \, , \quad p_{\hat{2}} =  p_2 + z q_2 \, , \quad 
p_{\hat{3}} =  p_3 + z q_3 \, ,
\eea
which means we perform the soft-BCFW shift on the dilaton (with $a_1=1$) and the usual BCFW shifts on two other legs. 
To preserve the on-shell conditions and the momenta conservation, we require $q_2$ and $q_3$ to satisfy
\bea
q_2^2=q_3^2=0 \,, \quad q_2 \cdot p_2= q_3 \cdot p_3 =0 \, , \quad q_2 + q_3 = p_1 \, ,
\eea
and we are allowed to choose any convenient solutions. 
Note from $q_2 + q_3 = p_1$, we have $q_2 \cdot p_1=0, \, q_3 \cdot p_1=0, \, q_2 \cdot q_3=0$ since they are all null momenta. 
The contour integral for deriving recursion relations becomes  
\bea
A_n(0) ={1 \over 2 \pi i} \oint_{|z|=0} dz {A_n(z) \over z(1-z)} \,, 
\eea
where a simple factor $(1-z)$ is enough to improve the large-$z$ behaviour here. 
From the residue theorem, we have
\bea \label{eq:recursion3}
A_n(0) = \sum_I{}^{'}  {1 \over P_I^2 - m_I^2} {A_L(z^-_I)A_R(z^-_I) \over (1 - z_I^-/z_I^+)(1-z_I^-)} +(z_I^- \leftrightarrow z_I^+) 
-
\sum^0_{k=-1} R^{(k)}_1  \, .
\eea
Similar with the discussion under (\ref{eq:recursion_general}), the ``primed sum" $\sum_I{}^{'}$ corresponds to factorization poles, but excludes the channels formed by three-point sub-amplitudes with leg $1$ on one side. 
Again, the reason is that such factorization poles contain ${1 \over 1-z}$, which has been included in the soft limit contribution $R^{(k)}_1$. 
As before, this subtlety can be discovered from the propagators of these factorization channels. 
For three-point amplitude $A_3(\hat{1}, i, \hat{P}_I)$ with $i \neq 2,3$, we have
\bea
{1 \over s_{\hat{1} i} - m_i^2} = {1 \over 2 p_{\hat{1}} \cdot p_i } = {1 \over 2 (1-z) p_1 \cdot p_i } \, .
\eea
While for the three-point factorization poles with $i$ being the shifted legs $2$ or $3$, we have
\bea
{1 \over s_{\hat{1} \hat{i} } - m_i^2} = {1 \over 2 p_{\hat{1}} \cdot p_{\hat{i}} } = 
{1 \over (1-z) } {1 \over 2  ( p_1 \cdot p_i + z p_1 \cdot q_i ) } \,.
\eea
Note $p_1 \cdot q_i=0$, thus they also  only contain the pole at $z=1$. 
Let us now consider the $R^{(k)}_1$ term. From the soft theorems, it is straightforward to see that $R^{(k)}_1$ are given by,
\bea
R^{(-1)}_1 = {1 \over 2 \pi i} \oint_{z=1} dz { \left(\mathcal{S}_{{\rm M}, 1}^{(-1)} A_{n-1,1}\right)(z) \over z (1-z)^2}
\, , \quad
R^{(0)}_1 = {1 \over 2 \pi i} \oint_{z=1} dz{\left[ ( \mathcal{S}_{{\rm M}, 1}^{(0)} + \mathcal{S}_1^{(0)} )A_{n-1,1}\right](z) \over z (1-z)} \, .
\eea
The first term $R^{(-1)}_1$ contains a double pole, which requires an expansion of $ \left(\mathcal{S}_{{\rm M}, 1}^{(-1)} A_{n-1}\right)(z)$ to order $\mathcal{O}(z-1)$, while the second term $R^{(0)}_1$ can be trivially computed out by setting $z=1$. 
Note that here the subleading terms in the soft theorems, $( \mathcal{S}_{{\rm M}, 1}^{(1)} + \mathcal{S}^{(1)}_1 )A_{n-1, 1}$, are actually not required. 
On the other hand, to compute amplitudes with only $m$ dilatons, the input of the lower-point amplitudes should be with $(m{-}1)$ dilatons. 
Thus, eventually the amplitudes of zero number of dilaton are needed. 
Apparently the soft theorems cannot provide useful information for the amplitudes without any dilaton, these amplitudes have to be computed by other means. 
For this particular case, the amplitudes $A_n(\phi)$ with massive scalar $\phi$ only, can be recursively determined by the usual BCFW recursion relations, if we wish. 
Since $A_n(\phi)$ only receives contributions from factorization diagrams, the amplitudes have a good large-$z$ behaviour if we perform $(n{-}1)$-line BCFW shifts, 
\bea
p_{\hat{i}} = p_i + z q_i \, , \quad {\rm for} \quad i=1, \ldots, n-1 \, ,
\eea
with $q_i^2=0$, $p_i \cdot q_i =0$ and $\sum^{n-1}_{i=1} q_i =0$. 

Let us now consider a few concrete examples. 
Here we take the dimension $d=4$ for the simplicity. 
Begin with the four-point amplitude $A(\xi_1, \xi_2, \phi_3, \phi_4)$, whose factorization channels contain only $1/(1-z)$ pole and are all already included in the soft terms $R^{(k)}_1$. 
From the contribution of $R^{(k)}_1$, we obtain
\bea \label{eq:4pts} 
R_1 &=& R^{(-1)}_1 + R^{(0)}_1
\cr
&=& -2i\lambda^2{1 \over 2 \pi i} \oint_{z=1} dz { 1 \over z (1-z)^2}  \left ( {m^2 \over p_1 \cdot ( p_3 + z q_3 ) } 
+ {m^2 \over p_1 \cdot p_4 }   \right) 
-
2i\lambda^2  {1 \over 2 \pi i}\oint_{z=1} dz{ 1 \over z (1-z)} \cr  
&=& 2i\lambda^2\left( {m^2 \over p_1 \cdot p_3 } + {m^2 \over p_1 \cdot p_4 }   \right) + 2i\lambda^2 \, , \eea
here again we have used the condition $p_1 \cdot q_3 =0$. 
This is exactly the result of four-point amplitude $A(\xi_1, \xi_2, \phi_3, \phi_4)$, which can be also computed from Feynman diagrams. 
One can also apply the recursion relations to higher-point amplitudes straightforwardly. 
For instance at five points, we have two different kinds of amplitudes: $A_{5}(\xi, \phi, \phi, \phi, \phi)$ and $A_{5}(\xi, \xi, \xi, \phi, \phi)$. 
To compute $A_{5}(\xi, \phi, \phi, \phi, \phi)$, we need the lower-point input $A_{4}(\phi, \phi, \phi, \phi)$, which is given by
\bea
A_{4}(\phi_1, \phi_2, \phi_3, \phi_4)=-4i\lambda^2\left({m^2 \over s_{12}} + {m^2 \over s_{13}}+ {m^2 \over s_{14}}\right) \, .
\eea
To calculate $A_{5}(\xi, \xi, \xi, \phi, \phi)$, we need $A_{4}(\xi, \xi, \phi, \phi)$ which has been computed previously in (\ref{eq:4pts}) by recursion relations. 
We have calculated both these two five-point amplitudes using the recursion relations and checked that they agree with the results from Feynman diagrams. 

\section{Conclusions} \label{section:conclusion}

In this paper, we derive new on-shell recursion relations for scattering amplitudes which have bad large-$z$ behaviour but satisfy certain soft theorems. 
This work is an generalization of the ideas in~\cite{Cheung:2015ota}. 
The crucial observation is that soft theorems provide additional information which can help to overcome the bad large-$z$ behaviour. 
First of all, it is straightforward to see that the recursion relations of~\cite{Cheung:2015ota} can apply to amplitudes containing particle species other than scalars, as far as the amplitudes also have the enhanced soft limits. 
In particular we apply the recursion relations to the amplitudes of the Akulov-Volkov theory of Goldstino's. 
So we show that the amplitudes of the Akulov-Volkov theory can be recursively constructed from four-point amplitudes, which are in fact completely fixed by the symmetries. 

We further generalize the recursion relations to amplitudes that do not vanish in the single-soft limit. 
As we emphasized that the construction of amplitudes having vanishing soft limits can be treated as a special case of those with non-vanishing soft limits. 
This general case can be further separated into two different situations. 
One is that amplitudes contain only Goldstone bosons. 
To utilize the soft theorems to derive the recursion relations, we perform the same soft-BCFW shifts as done in~\cite{Cheung:2015ota}. 
The other more general situation is that amplitudes contain both Goldstone bosons and non-Goldstone particles. Then we perform mixed shifts with soft-BCFW shifts on some Goldstone bosons and the usual BCFW shifts on some other legs, which can be Goldstone bosons or not. 
The result of the recursion relations contains the usual factorization terms, as well as contributions from the non-vanishing soft limits which are determined by the soft theorems. 
The recursion relations provide a proof: those tree-level amplitudes satisfying ``good-enough" soft theorems are on-shell constructible. 
We apply our new recursion relations to amplitudes consisting of dilatons in effective field theories with spontaneously-broken conformal symmetries, include the dilation effective actions~\cite{Komargodski:2011vj, Elvang:2012st, Elvang:2012yc} as well as a two-scalar model~\cite{Boels:2015pta}. 

It may be not so surprised that the recursion relations work for the amplitudes in the Akulov-Volkov theory, since after all they are supersymmetricly related to the scalar DBI theory~\cite{Chen:2015hpa} (although it should be made clear that SUSY Ward identity is not enough to fix one from the other~\cite{Elvang:2009wd} at high enough multiplicities, in particular beyond six points). 
However, it turns out that the same recursion relations cannot be applied to gluon amplitudes in the BI theory, which supposes to be related to the scalar DBI theory via supersymmetry as well. 
Indeed it is easy to check that the soft behaviour is not enough to overcome the large-$z$ growth of amplitudes in the BI theory. 
As we have discussed in section \ref{section:fermions}, it would be very interesting to explore the supersymmetric extension of such recursion relations. 

Another interesting direction would be to find more examples whose amplitudes have bad large-$z$ behaviour, but the recursion relations proposed in this paper can still apply.
For instance, one interesting case would be gravity theories with higher-derivative terms. 
One interesting question is that what kind of higher-derivative terms can be added to the Einstein theory such that recursion relations in this paper are still available.  
Apparently, higher-derivative terms will lead to a bad large-$z$ behaviour. 
At the same time, terms like $\phi R^2$ would even destroy the universality of the sub-sub-leading order~\cite{Bianchi:2014gla} in the soft graviton theorems~\cite{Gross:1968in, Cachazo:2014fwa}. 
Finally, in this paper, we have only used the information of the single-soft limit, which is enough for the cases considered in this paper. 
It is known that amplitudes in many theories also enjoy universal behaviours in the limits with more than one leg being soft, see e.g. references~\cite{ArkaniHamed:2008gz,Chen:2014xoa,Chen:2014cuc, Cachazo:2015ksa, Klose:2015xoa, Volovich:2015yoa, Du:2015esa, DiVecchia:2015bfa, Low:2015ogb}. 
It would be of great interest to explore the applications of soft theorems with multiple-soft legs in the context of the recursion relations. 

\section*{Acknowledgements}
We would like to thank Massimo Bianchi, Rutger Boels, Raffaele Marotta, Jaroslav Trnka and Yang Zhang for helpful discussions. H.L. is supported by the German
Science Foundation (DFG) under the Collaborative Research Center (SFB) 676 ``Particles, Strings and the Early Universe".

\bibliographystyle{JHEP}
\bibliography{Refs}

\end{document}